\documentclass[aps,prl,twocolumn,a4paper,10pt,notitlepage,footinbib,superscriptaddress,longbibliography]{revtex4-1}
\usepackage[english]{babel}
\usepackage{lmodern}
\usepackage[utf8]{inputenc}
\usepackage{endnotes}
\usepackage{amssymb,amsmath,amsfonts}
\usepackage{textcomp}
\usepackage{braket}
\usepackage{ulem}
\usepackage{soul}
\usepackage{comment}
\usepackage{graphicx}
\usepackage{dcolumn}
\usepackage{bm}
\usepackage{xcolor}

\begin{document}

\title{Negative quantum friction in nanoscale water flows: the Wigner picture}

\author{Adriano Tiribocchi}
\affiliation{Istituto per le Applicazioni del Calcolo, Consiglio Nazionale delle Ricerche, Via dei Taurini 19, Rome, I-00185, Italy}
\affiliation{INFN Tor Vergata, Via della ricerca scientifica 1, Rome, I-00133, Italy}

\author{Marco Lauricella}
\affiliation{Istituto per le Applicazioni del Calcolo, Consiglio Nazionale delle Ricerche, Via dei Taurini 19, Rome, I-00185, Italy}

\author{Efthimios Kaxiras}
\affiliation{Department of Physics, Harvard University, 17 Oxford St, Cambridge, MA 02138, United States}
\affiliation{School of Engineering \& Applied Sciences, Harvard University, Cambridge, Massachusetts 02138, USA}

\author{Sauro Succi}
\affiliation{Istituto per le Applicazioni del Calcolo, Consiglio Nazionale delle Ricerche, Via dei Taurini 19, Rome, I-00185, Italy}
\affiliation{Center for Life Nano- \& Neuro-Science, Fondazione Istituto Italiano di Tecnologia, viale Regina Elena 295, 00161
Rome, Italy}
\affiliation{Department of Physics, Harvard University, 17 Oxford St, Cambridge, MA 02138, United States}

\begin{abstract}
We explore the phenomenon of "quantum" friction based on
a single-particle model patterned after the Wigner equation
describing electrons flow in a solid wall confining nanoscale
water flows.
The numerical simulations show a clear signature of negative 
quantum friction, namely a net momentum transfer from the electrons
in the solid wall to the flowing water molecules.
Such net momentum transfer results into a sizeable reduction of the
water friction, up to forty percent, depending on the strength
of the coupling between classical and quantum fluctuations. Our results offer the prospect of a theoretical framework bridging classical and quantum description by using continuum kinetic theories and particle-based simulations. 

\end{abstract}

\maketitle

\begin{figure*}[htbp]
\includegraphics[width=1.0\textwidth]{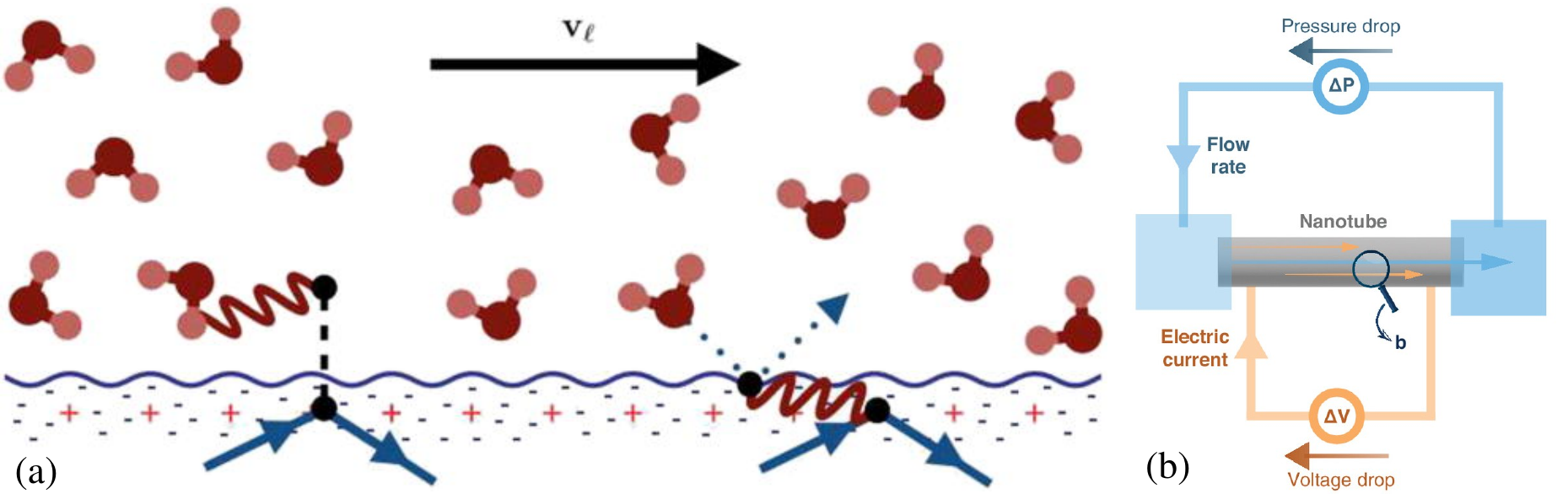}
\caption{(a) Schematic representation of the mechanisms driving the electronic 
current: hydrons (charged fluctuations of water molecules, red particles) transfer 
momentum to the electrons in the solid (black particles) through screened Coulomb 
interactions (black dotted line) and phonons (wavy lines), resulting from hits against the wall. (b) 
Schematic functioning of a hydronic generator. 
As the liquid flows through a nanotube connected to an external circuit, an 
electric current,  caused by the hydroelectronic drag, results 
from the flow of the liquid.
Figures adapted from Refs.\cite{boq_prx,boq_pnas}.}\label{fig1}
\end{figure*}

%\section{Introduction}
In recent years, nanofluidic experiments have shown ample evidence that channel walls significantly impact the dynamics and the transport properties of the confined fluid, supporting the view that non-trivial interactions localized at the fluid-solid interface stem from the coupling between the electronic degrees of freedom of the wall and the liquid  \cite{Schoch2008,bocquet2014,kavokine2021,sutter2025}.  These results critically challenge the paradigm that the channel wall is a passive boundary condition, pointing to a picture where nanochannels are actually hybrid quantum-classical systems whose properties shape fluid transport \cite{boq_nat}.
In this respect, it has recently been pointed out that the coupling  of charge fluctuations in nanoscale water flows and the electrons
in the confining solid (typically carbon nanotubes)
can give rise to a net momentum transfer between the liquid and the
solid, a phenomenon dubbed "quantum friction" \cite{boq_nat,boq_prx,boq_pnas,kavo,boq_nat_comm}. 
Quantum friction originates from a general fluctuation-induced dissipation mechanism that can be driven by either classical thermal fluctuations or genuinely quantum fluctuations. However, while the underlying momentum-transfer process does not rely on quantum mechanics {\it per se}, 
quantum effects become essential when the relevant fluctuations involve high-frequency modes ($\hbar\omega\gtrsim k_B T$), where Fermi statistics and quantum correlations cannot be neglected. Quantum friction can take both signs, meaning that it can reduce
the drag experienced by the flowing water, or produce electric currents
by directly driving the electrons on the solid (quantum current drive). The idea is sketched in Fig.\ref{fig1}a where water molecules, driven by an external 
pressure gradient, are subject to nanoscale charge fluctuations (termed "hydrons") and 
transfer momentum to the electron in the solid through screened Coulomb interaction. 
Concurrently, collisions of the water molecules with the solid walls generate 
a "phonon" wind that eventually drags the electrons along. This fluctuation-induced coupling between fluid and electron transport could be at the root of promising practical applications, such as the possibility  of converting the kinetic energy of a small scale flow into an electronic current 
by leveraging the hydroelectric drag, in which fluid ions induce an electronic current through the tube wall of the nanochannel (Fig.\ref{fig1}b and \cite{boq_pnas}). 

In  this letter we study, by numerical simulations, the phenomenon of quantum friction by modeling electron dynamics via a fractional Langevin equation of a single particle patterned from the Wigner equation of motion of electrons flowing in the walls of a nanochannel. The bridge between the continuum kinetic description and the particle-based dynamics is built by incorporating,
into the Langevin equation, 
classical and quantum force fields, where the latter stem from a hierarchical expansion of high order derivatives in the momentum space of the Wigner equation and are integrated on a fractional timestep with exponent resulting from dimensional analysis.

For sufficiently high values of quantum fluctuations, we find a significant reduction of the electron current, which we associate with a 
signature of negative quantum friction, where e net momentum has been transferred to the water molecules. Depending upon the magnitude of such fluctuations, current reduction attains values up to $40\%$ lower than the classical counterpart and in satisfactory agreement with the theoretical results of Ref.\cite{boq_prx}, which are based on the Keldysh formalism of non-equilibrium quantum statistical mechanics
\cite{keldysh,Kadanoff,Rammer}.

Quantum friction was first analyzed and predicted within such framework,
which 
captures fluctuation-induced interactions and frictional forces that are typically  absent from mean field treatments, such as the Poisson-Nerst-Planck-Stokes transport approach \cite{andelman1995,kavokine2021}.
The Keldysh formalism is essentially built starting from a two-body Green's function technique
\begin{equation}
G(1,2) \equiv \langle \Phi(1) \Phi(2) \rangle,
\end{equation}
where $\Phi$ is the many-body quantum wavefunction and 
$1 \equiv (\vec{x}_1,t_1)$ and $2 \equiv (\vec{x}_2,t_2)$ 
are shorthands notation for spacetime coordinates of each particle.
The formalism is very intensive and, under appropriate simplifications (i.e. homogeneity), lends itself to analytical calculations, which is 
precisely the route that has revealed the phenomenon of quantum friction \cite{boq_nat,boq_prx}.
Although this route is fundamental and very insightful, the homogeneity assumptions
cast questions on its viability and accuracy for situations of experimental
relevance where those simplified assumptions are typically broken \cite{sutter2025}. These limitations provide the motivation towards alternative representations
of quantum transport phenomena, such as the Wigner representation pursued in the present work.

The Wigner distribution is defined by the following transformation \cite{wigner,HILLERY1984121}
\begin{equation}
\mathcal{W}(\vec{x},\vec{p},t,E) = 
\int \Phi(\vec{x}-\vec{r},t-z)
e^{i \frac{\vec{p} \cdot \vec{r} + E s}{\hbar}} 
\Phi(\vec{x}+\vec{r},t+s) \; d\vec{r} ds
\end{equation}
where
$\vec{x}_{1,2} = \vec{x} \pm \vec{r}$, $t_{1,2} = t \pm s$, $E$ is the 
particle energy, $\vec{p}$ is  the momentum and $\hbar$ is the reduced Planck constant. 
To this point, the Wigner distribution is fully equivalent to the
two-body Green's function, as they both depend on eight independent 
coordinates. A major merit of the Wigner formulation is that
it casts the quantum problem into a classical-looking
phase-space kinetic equation, thereby facilitating the
analysis of the transition from quantum to classical behavior \cite{polkovnikov}.

Such transition is controlled by two main assumptions.
First, quantum excitations should be weakly interacting, meaning
that there is a one-to-one correspondence between the
particle energy and its momentum, $E=E(\vec{p})$, the so called
dispersion relation. Under such an assumption, one can write
$\mathcal{W}(\vec{x},\vec{p},E,t) = 
W(\vec{r},\vec{p},t)\delta(E-E(\vec{p}))$
and show that the reduced Wigner distribution obeys a Boltzmann-like
kinetic equation
\begin{equation}\label{wig_eq}
\partial_t W + p \partial_x W + F_1 \partial_p W +
\sum_{l=1} \left(\frac{\hbar}{2i}\right)^{2l} F_{2l+1} \partial_p^{2l+1} W = Q,  
\end{equation} 
where  $F_{2l+1} \equiv \partial_x^{2l+1} V$,
$V$ being the interaction potential, and 
$Q$ is a collision term resulting from the scattering processes between 
the quasi-particles \cite{succi1,succi2,succi3}. 
In Eq.(\ref{wig_eq}) vector notation has been relaxed 
and the particle mass has been set to unity for the sake of simplicity.

A few comments are in order.
First, the term under summation describes the effects
of quantum interference through a hierarchical expansion
in higher order derivatives in momentum space, coupled
to correspondingly higher order spatial derivatives of the
distribution function. In other words, quantum fluctuations probe 
increasingly non-local features of the interaction potential.
Note that the "force" terms $F_{2l+1}$ have dimension
$p^{2l+1}/t$, hence they represent higher order dynamical processes, a
point to which we shall return later.
Denoting by $\delta$ the typical lengthscale
of the potential and $p_T = m v_T$ the typical thermal momentum,
each term of the series scales like $q^{2l}$ compared to the classical
forcing $F_1 \partial_p W$, where
\begin{equation}
q \equiv \frac{\lambda_T}{\delta}
\end{equation}      
and $\lambda_T = \hbar/p_T$ is the thermal De Broglie length. For the third order term, for example, we have $F_3\partial_p^3W/F_1\partial_pW\sim q^2$.
Hence, the parameter $q$ is a measure of the "quantumness"
of the problem.
The limit $q=0$ stands for a fully classical case, whereas $q \ge 1$ describes a quantum problem, for which all terms
of the hierarchy need to be accounted, fully exposing the non-local
nature of the Wigner equation.   
In this paper we shall be concerned with both the quasi-classical regime,
$q^2 \le 0.01$, and semi-quantum regimes $0.01 < q^2 < 0.05$.
As a result, we shall confine our attention to the leading
term $F_3$ in the quantum fluctuating component of the Wigner equation.
 
This term has dimensions $p^3/t$, hence it 
describes "fractional" motion with non integer exponent $1/3$
$$
dp^3/dt = F_3 \leftrightarrow dp = F_3^{1/3} dt^{1/3}.
$$ 
The above equivalence is not rigorous, but it stands as a {\it bona-fide} 
model of fractional motion for a particle obeying the following dynamics

\begin{eqnarray}
dx &=& p dt\label{frac_lang1}\\   dp &=& -\gamma (p-p_{\phi}) dt - F_1 dt + F_{1/3} dt^{1/3},\label{frac_lang2}
\end{eqnarray}
where $p_{\phi}$ is the phonon wind acting as a source of momentum for the
electrons in the solid and $\gamma$ is the classical friction.
This is a fractional Langevin-like equation where, besides the systematic contribution 
multiplied by $dt$, one also has a high-order "forcing" term multiplied by $dt^{1/3}$ \cite{weiss}. The numerical integration of Eq.(\ref{frac_lang1})
 and Eq.(\ref{frac_lang2}) proceeds via a standard velocity Verlet algorithm where  where $F_1=\partial_x V$ and 
\begin{equation}\label{F3}
F_{1/3}(x(t)) = \frac{1}{\Gamma(1/3)} \int_0^t (t-t')^{-2/3} F_3^{1/3}(x(t')) dt',
\end{equation}
reflecting the non-local
nature of fractional derivatives.
In Eq.(\ref{F3}), $F_{1/3}(x)$ is expressed as a Riemann-Liouville fractional integral of order $1/3$ (turning $F_3$ into a force with power-law memory), $\Gamma$ is  the Gamma function and $F_3=\partial^3_x V$, where the numerical derivatives are computed using standard finite difference schemes.

We consider a particle moving along the solid wall confining a 
nanoscale water channel, modeled as a square domain of linear size $L=128$, where the main force experienced by the particle is the
stochastic potential $V(x)$ associated with charge fluctuations of the water molecules. The potential  experienced
by an electron located at position $x$ in the solid substrate is given by
\begin{equation}
V(x) = \int_{0}^L \int_{0}^L 
\frac{\tilde z(\xi, \eta)}
{[(\xi-x)^2 + \eta^2]^{1/2}} \; d \xi d\eta, 
\end{equation}
where $\tilde z(\xi,\eta)$ is the 
charge density, $<\tilde z >=0$ and $< \tilde z^2> = W$. 
Since $V(x)$ is expected to vary much slower than the particle dynamics, it is updated on a slower timescale (typically every $10$ time steps) than the particle dynamics, thus implementing a quasi-static adiabatic approximation. Although other formulations could be, in principle, used, the one adopted here ensures good numerical stability and reduced computational cost.
Also, at the boundaries 
the potential is treated as effectively periodic via buffer extension, while the forces are set to zero to ensure numerical stability in the evaluation of finite-difference derivatives.
The fractional particle motion is integrated for a series of values 
of the fluctuation amplitude $W$,  ranging from $0.01$ to $0.4$, and quantumness  $q^2$ between $0$ (classical regime) and $0.05$ (semi-quantum regime). 
Moreover, in our simulations  we fix $p_{\phi}=1$, $\gamma=0.1$ and the timestep  $dt=0.1$.

\begin{figure}[htbp]
\centering
\includegraphics[width=1.0\columnwidth]{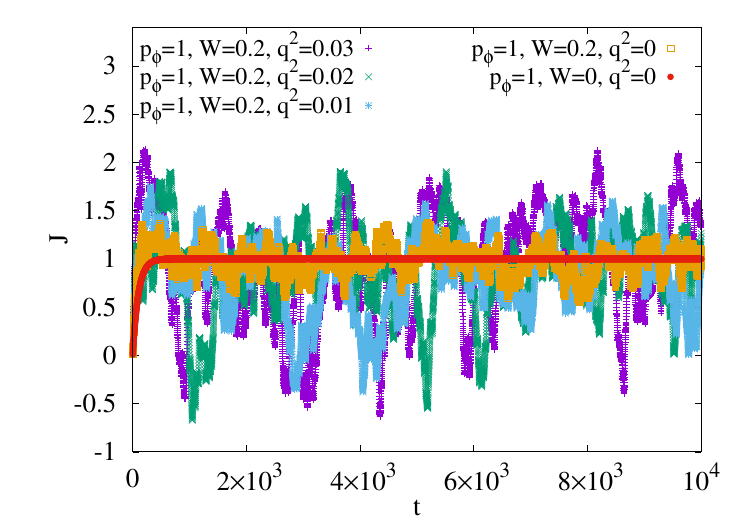}
\caption{Time evolution of the current $J$ of a single particle 
(extracted from $N=100$ different realizations) 
for $p_{\phi}=1$, $W=0,0.2$ and $q^2=0,0.01,0.02,0.03$.
The solid curve represents the deterministic case, without water 
fluctuations, whereby the particle is driven towards the steady
state condition $p=p_{\phi}$ on a time scale $1/\gamma$ (classical friction). Increasing $q^2$ leads to larger amplitude fluctuations and a slight decrease of the average current.
}\label{fig2}
\end{figure}

\begin{figure}[htbp]
\centering
\includegraphics[width=1.0\columnwidth]{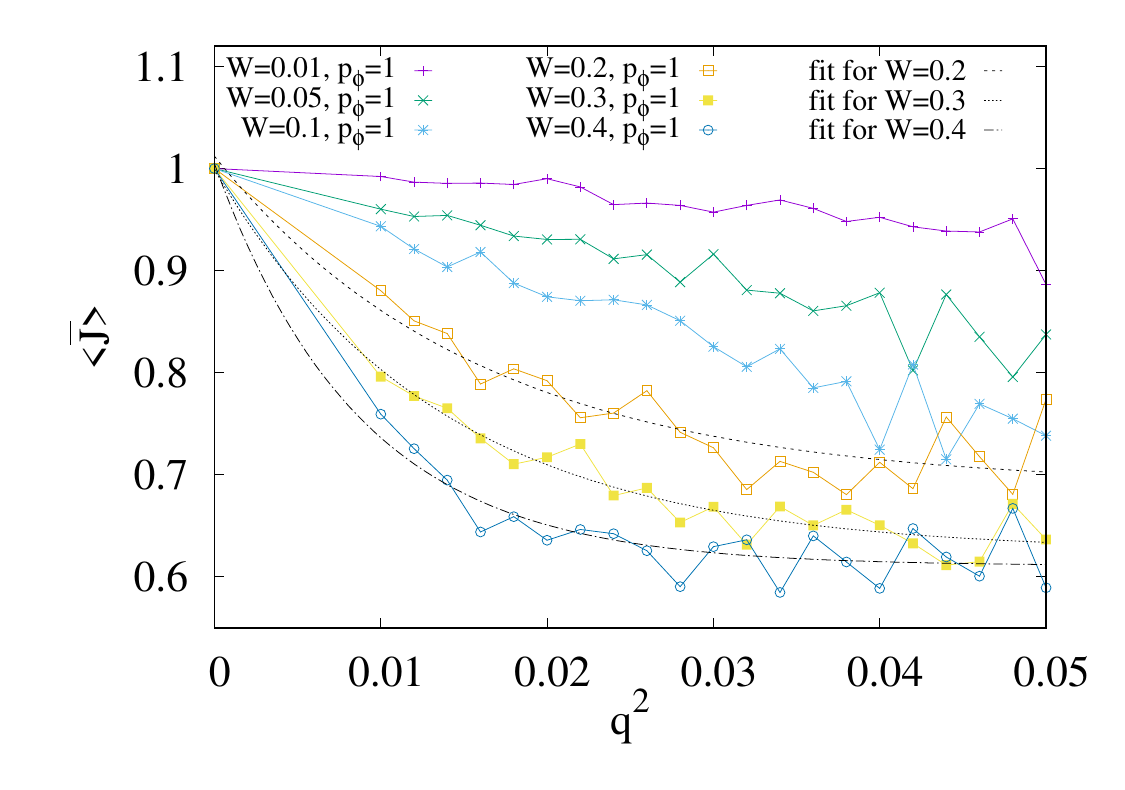}
\caption{Current {\it vs} $q^2$, averaged over time ($T=20,000$) and 
realizations ($N=100$) for different values of $W$ and $p_{\phi}=1$.
A net decrease of the electronic current at increasing $q^2$ is visible, 
amounting to about $30$ percent at $q^2=0.05$ (semi-quantum) regime.
The curve obeys a fit of the form
$J(q^2)=J(0) - \Delta J (1-exp(-q^2/Q^2))$, with $\Delta J/J(0) \sim 0.32$
and $Q^2 \sim 0.015$ for W=0.2 (dashed curve), $\Delta J/J(0) \sim 0.37$ and $Q^2\sim 0.013$ for 
$W=0.3$ (dotted curve), $\Delta J/J(0) \sim 0.4$ and $Q^2\sim 0.009$ for $W=0.4$ (dot-dashed curve).}\label{fig3}
\end{figure}

In Fig.\ref{fig2} we show the typical time evolution of the current $J$ (i.e. the particle momentum) of a single particle for the reference case $p_{\phi}=1$ and $W=0$ (i.e. in the absence of charge fluctuations) and for different values of $q^2$ (with $W=0.2$). 
The classical stochastic motion (i.e. $q^2=0$ and W$=0.2$) performs sizeable excursions
around the deterministic ($W=0$) value $p_{\phi}=1$, as dictated by the structure of the potential $V(x)$, where $W$ sets the amplitude of the fluctuations around $p_{\phi}$ (see SI for further details).  
By switching on the quantum "forces" (and keeping $W$ constant), positive and negative excursions appear, whose amplitude increases with $q^2$. This would initially suggest that the quantumness induces a change of amplitude of the current without a net effect on its average. However,
by averaging over ensemble and time
we observe a net reduction of the electron current $\langle\bar{J}\rangle$ at increasing values of $q^2$. In Fig.\ref{fig3} 
we report evidence of this behavior and show that $\langle\bar{J}\rangle$ diminishes with increasing  $q^2$ and $W$. At low values of $W$, the decrease is mild and stabilizes approximately around $5\%$ of the unperturbed value while, at increasing $W$, it attains almost  $40\%$ of the unperturbed current for $W=0.4$ and $q^2=0.05$, a result in agreement with the ones presented in Ref.\cite{boq_prx}. This is  precisely the negative quantum friction scenario, where  the electrons return part of the momentum (hence they slow down) imparted by the phonon wind back to the water molecules. Inspecting the current profiles, we find that they follow an exponential decay of the form $J(q^2)=J(0)-\Delta J(1-exp(-q^2/Q^2))$, where the relaxation time $Q^2$ decreases with increasing $q^2$ and the ratio $\Delta J/J(0)$ accounts for the current reduction.

\begin{figure*}[htbp]
\centering
\includegraphics[width=1.0\textwidth]{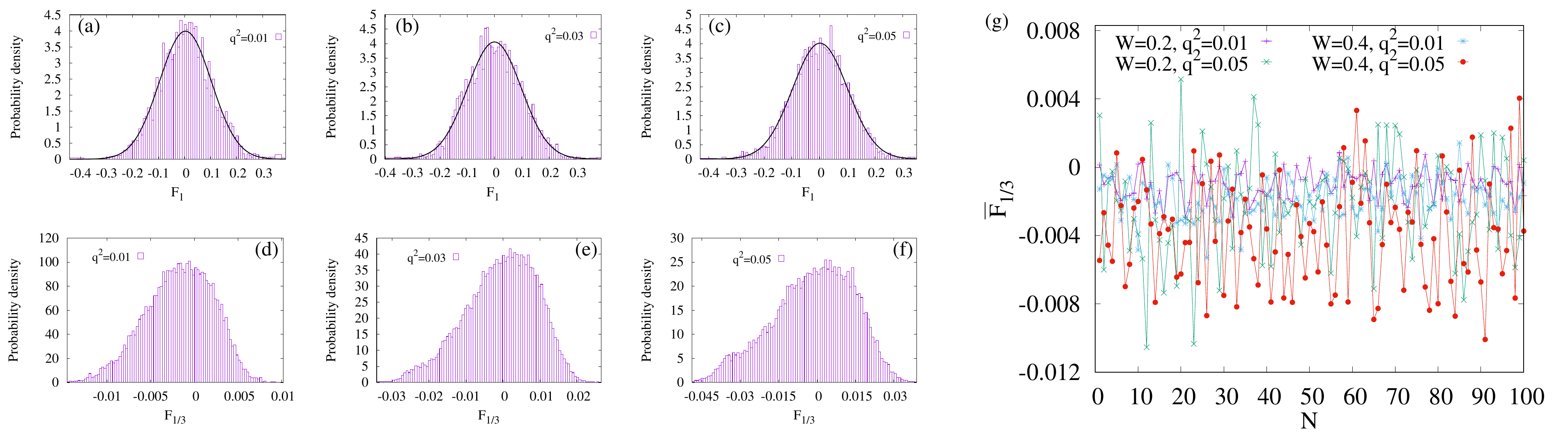}
\caption{The probability density of $F_1$ (a,b,c) and $F_{1/3}$ (d,e,f) for $p_{\phi}=0.1$, $W=0.2$  and  
$q^2=0.01$, $q^2=0.03$, $q^2=0.05$.
Regardless of the values of $q^2$, $F_1$ closely follows a Gaussian distribution where $\mu\sim  0$ and $\sigma\sim 0.1$. By contrast, $F_{1/3}$ shows a negative skewness, with an increasingly heavy left tail, as $q^2$ augments, with the skewness approximately ranging from -$0.3$ for $q^2=0.01$ to $-0.4$ for $q^2=0.05$.
We associate the progressive increase of negative skewness with the reduction of $\langle\bar{J}\rangle$.
In panel (g) we show the typical trend of $\bar{F}_{1/3}$ over $N=100$ realizations. Increasing $q^2$ leads to a shift towards negative values with a concurrent increase of the fluctuation amplitude.}\label{fig4}
\end{figure*}

By inspecting the instantaneous force profiles, it is apparent that $F_1(x)$ show Gaussian fluctuations
around zero for all values of $q^2$ (Fig.\ref{fig4}a,b,c),
while $F_{1/3}(x)$ exhibits a negative skewness, whose left tail becomes heavier for increasing $q^2$ (Fig.\ref{fig4}d,e,f). This essentially results from the structure of the kernel of $F_{1/3}$, which  heavily weights recent events while retaining long-memory contributions. In the presence of rapid sign alternation, this induces incomplete cancellation between positive and negative fluctuations, leading to a systematic imbalance in higher-order moments and effectively breaking the statistical symmetry between positive and negative states. 
Sudden changes of sign are indeed observed for $F_3$, which exhibits a bimodal distribution with a spreading that augments for increasing $q^2$ (see SI for further details). The latter is because 
$F_3$ experiences much sharper transitions 
between negative and positive values, thus its statistics around zero (essentially at the boundaries where $F_1=0$ and $F_3=0$) remains largely unpopulated. The time average $\bar{F}_{1/3}$ for different values of $N$ also shows that increasing $q^2$ induces larger fluctuations with a net shift towards negative values, further corroborating the decrease of electron current (averaged over time and realizations) shown in Fig.\ref{fig3}.

The present results indicate 
that the coupling between quantum fluctuations and classical 
heterogeneity can lead to a sizeable reduction of the friction experienced  
by the nanoscale water flow, due to a net momentum transfer 
from the electrons back to the water molecules.
This provides an additional and independent evidence of genuinely quantum friction effects in nanoscale water flows. On the modeling side, our approach provides a viable strategy to  capture such effects as an alternative to {\it ab-initio} molecular dynamic simulations which, despite being an illuminating guide in  revealing the molecular and electronic structure of liquid–solid interfaces \cite{ma2011,tocci2014,ma2015,askerka2016,brandenburg2019,schran2021},
have been sofar unsuccessful in simulating quantum friction and hydro-electronic drag. 
Future work will include the effect of classical heterogeneity, such as
nano-patterns and geometrical/chemical coating \cite{papa2019}, on the mechanism of quantum
friction, possibly leading to the optimal design of a new class
quantum-engineered nano-fluidic devices for energy conversion applications.

\section{Acknowledgments} 
SS is grateful to L. Bocquet for introducing him to the subject
of quantum nanofluidics and to N. Kavokine and B. Coquinot for many valuable discussions.  
This work was triggered by the Symposium "La nanofluidique à la crois\'ee des chemins",
Collège de France, Paris, May 2023.

\bibliography{biblio} 

@BOOK{weiss,
	title={Quantum dissipative systems},
	publisher={World Scientific Publishing},
	year={2021},
	author={Weiss, U.},
    DOI={https://doi.org/10.1142/12402},
}

@BOOK{Rammer,
	title={Quantum Field Theory of Non-Equilibrium States},
	publisher={Cambridge University Press},
	year={2007},
	author={Rammer, J.},
}

@BOOK{Kadanoff,
	title={Quantum Statistical Mechanics: {G}reen's Function Methods in Equilibrium and Nonequilibrium Problems},
	publisher={W. A. Benjamin},
	year={1962},
	author={Kadanoff, L. P. and Baym. G.},
}

@article{brandenburg2019,
author = {Brandenburg, J. G. and Zen, A. and Alfè. D. and Michaelides, A.},
title = {Interaction between water and carbon nanostructures: {H}ow good are current density functional approximations?},
journal = {J. Chem. Phys.},
volume = {151},
pages = {164702},
year = {2019},
doi = {https://doi.org/10.1063/1.5121370},
}

@article{askerka2016,
author = {Askerka, M. and Maurer, R. J. and Batista, V. S. and Tully, J. C.},
title = {Role of Tensorial Electronic Friction in Energy Transfer at Metal Surfaces},
journal = {Phys. Rev. Lett.},
volume = {116},
pages = {217601},
year = {2016},
doi = {https://doi.org/10.1103/PhysRevLett.116.217601},
}

@article{papa2019,
author = {Papadopoulou, E. and Megaridis, C. M. and Walther, J. H. and Koumoutsakos, P.},
title = {Ultrafast Propulsion of Water Nanodroplets on Patterned Graphene},
journal = {ACS Nano},
volume = {13},
pages = {5465-5472},
year = {2019},
doi = {https://doi.org/10.1021/acsnano.9b00252},
}

@article{tocci2014,
author = {Tocci, G. and Joly, L. and  Michaelides, A.},
title = {Friction of Water on Graphene and Hexagonal Boron Nitride from Ab Initio Methods: {V}ery Different Slippage Despite Very Similar Interface Structures},
journal = {Nano Letters},
volume = {14},
pages = {6872-6877},
year = {2014},
doi = {https://doi.org/10.1021/nl502837d},
}

@article{schran2021,
author = {Schran, C. and Thiemann, F. L. and Rowe, P. and M\"uller, E. A. and Marsalek, O. and Michaelides, A.},
title = {Machine learning potentials for complex aqueous systems made simple},
journal = {Proceedings of the National Academy of Science, USA},
volume = {118},
pages = {e2110077118},
year = {2021},
doi = {https://doi.org/10.1073/pnas.2110077118},
}

@article{ma2011,
author = {Ma, J. and Michaelides, A. and Alfè, D. and Schimka, L. and Kresse, G. and Wang, E.},
title = {Adsorption and diffusion of water on graphene from first principles},
journal = {Phys. Rev. B},
volume = {84},
pages = {033402},
year = {2011},
doi = {https://doi.org/10.1103/PhysRevB.84.033402},
}

@article{ma2015,
author = {Ma, M. and Grey, F. and Shen, L. and Urbakh, M. and Wu, S. and Liu, J. Z. and Zheng, Q.},
title = {Water transport inside carbon nanotubes mediated by phonon-induced oscillating friction},
journal = {Nature Nanotech.},
volume = {10},
pages = {692-695},
year = {2015},
doi = {https://doi.org/10.1038/nnano.2015.134},
}

@article{boq_pnas,
author = {Coquinot, B.  and Bocquet, L.  and Kavokine, N.},
title = {Hydroelectric energy conversion of waste flows through hydroelectronic drag},
journal = {Proceedings of the National Academy of Sciences},
volume = {121},
pages = {e2411613121},
year = {2024},
doi = {10.1073/pnas.2411613121},
}

@article{andelman1995,
author = {Andelman, D.},
title = { Electrostatic properties of membranes: {T}he {P}oisson-{B}oltzmann theory},
journal = {Handb. Biol. Phys.},
volume = {1},
pages = {603-642},
year = {1995},
doi = {https://doi.org/10.1016/S1383-8121(06)80005-9},
}

@article{boq_prx,
  title = {Quantum Feedback at the Solid-Liquid Interface: {F}low-Induced Electronic Current and Its Negative Contribution to Friction},
  author = {Coquinot, B. and Bocquet, L. and Kavokine, N.},
  journal = {Phys. Rev. X},
  volume = {13},
  pages = {011019},
  year = {2023},
  doi = {10.1103/PhysRevX.13.011019},
}

@article{sutter2025,
  title = {Fluctuation-induced and quantum effects in nanofluidic transport},
  author = {Sutter, A. and Gispert, P. and Coquinot, B. and Bocquet, L. and Kavokine, N.},
  journal = {arXiv:2606.06693},
  volume = {},
 pages = {},
  year = {},
  doi ={},
}

@article{Schoch2008,
  title = {Transport phenomena in nanofluidics},
  author = {Schoch, R. B. and Han, J. and Renaud, P.},
  journal = {Rev. Mod. Phys.},
  volume = {80},
  pages = {839--883},
  year = {2008},
  doi = {10.1103/RevModPhys.80.839},
}

@article{bocquet2014,
  title = {Physics and technological aspects of nanofluidics},
  author = {Bocquet, L. and Tabeling, P.},
  journal = {Lab Chip},
  volume = {14},
  pages = {3143-3158},
  year = {2014},
  doi = {https://doi.org/10.1039/C4LC00325J},
}

@article{kavokine2021,
  title = {Fluids at the Nanoscale: {F}rom Continuum to Subcontinuum Transport
},
  author = {Kavokine, N. and Netz, R. R. and Bocquet, L.},
  journal = {Annu. Rev. Fluid Mech.},
  volume = {53},
  pages = {377-410},
  year = {2021},
  doi = {https://doi.org/10.1146/annurev-fluid-071320-095958},
}

@article{polkovnikov,
  title = {Phase space representation of quantum dynamics},
  author = {Polkovnikov, A.},
  journal = {Annals of Physics},
  volume = {325},
  pages = {1790-1852},
  year = {2010},
  doi = {https://doi.org/10.1016/j.aop.2010.02.006},
}

@article{boq_nat,
  title = {Fluctuation-induced quantum friction in nanoscale water flows},
  author = {Kavokine, N. and Bocquet, M. L. and Boquet, L.},
  journal = {Nature},
  volume = {602},
  pages = {84-90},
  year = {2022},
  doi = {https://doi.org/10.1038/s41586-021-04284-7},
}

@article{wigner,
  title = {On the Quantum Correction For Thermodynamic Equilibrium},
  author = {Wigner, E.},
  journal = {Phys. Rev.},
  volume = {40},
  pages = {749-759},
  year = {1932},
  doi = {10.1103/PhysRev.40.749},
}

@article{kavo,
  title = {Strong Electronic Winds Blowing under Liquid Flows on Carbon Surfaces},
  author = {Liz\'ee, M. and Marcotte, A. and Coquinot, B. and Kavokine, N. and Sobnath, K. and Barraud, C. and Bhardwaj, A. and Radha, B. and Nigu\'es, A. and Bocquet, L. and Siria, A.},
  journal = {Phys. Rev. X},
  volume = {13},
  pages = {011020},
  year = {2023},
  doi = {10.1103/PhysRevX.13.011020},
}

@article{boq_nat_comm,
  title={Anomalous friction of supercooled glycerol on mica},
  author={Liz\'ee, M. and Coquinot,    B. and Mariette, G. and et al.},
   journal={Nat. Commun.},
  volume={15},
  pages={6129},
  year={2024},
  doi={https://doi.org/10.1038/s41467-024-50232-0},
}

@article{HILLERY1984121,
title = {Distribution functions in physics: Fundamentals},
journal = {Physics Reports},
volume = {106},
pages = {121-167},
year = {1984},
doi = {https://doi.org/10.1016/0370-1573(84)90160-1},
author = {Hillery, M.  and O'Connell, R. F. and Scully, M. O. and Wigner, E. P.},
}

@article{Succi1,
  title = {Lattice {W}igner equation},
  author = {Sol\'orzano, S. and Mendoza, M. and Succi, S. and Herrmann, H. J.},
  journal = {Phys. Rev. E},
  volume = {97},
  pages = {013308},
  year = {2018},
  doi = {10.1103/PhysRevE.97.013308},
}

@article{Succi2,
  title = {Keldysh Lattice {B}oltzmann Approach to Quantum Nanofluidics},
  author = {Succi, S. and Lauricella, M. and Montessori, A.},
  journal = {AIAA Journal},
  volume = {63},
  pages = {4},
  year = {2025},
  doi = {https://doi.org/10.2514/1.J064211},
}

@article{Succi3,
  title = {},
  author = {Succi, S. and Lauricella, M. and Tiribocchi, A.},
  journal = {Physics of Fluids},
  volume = {37},
  pages = {047122},
  year = {2025},
  doi = {https://doi.org/10.1063/5.0260872},
}

@article{keldysh,
  title = {Diagram Technique for Nonequilibrium Processes},
  author = {Keldysh, L. V.},
  journal = {Sov. Phys. JETP},
  volume = {20},
  pages = {1018},
  year = {1965},
  doi = {https://doi.org/10.1142/9789811279461_0007},
}
\bibliographystyle{unsrt}

\section{Supplemental Material}

\begin{figure*}[htbp]
\centering
\includegraphics[width=1.0\textwidth]{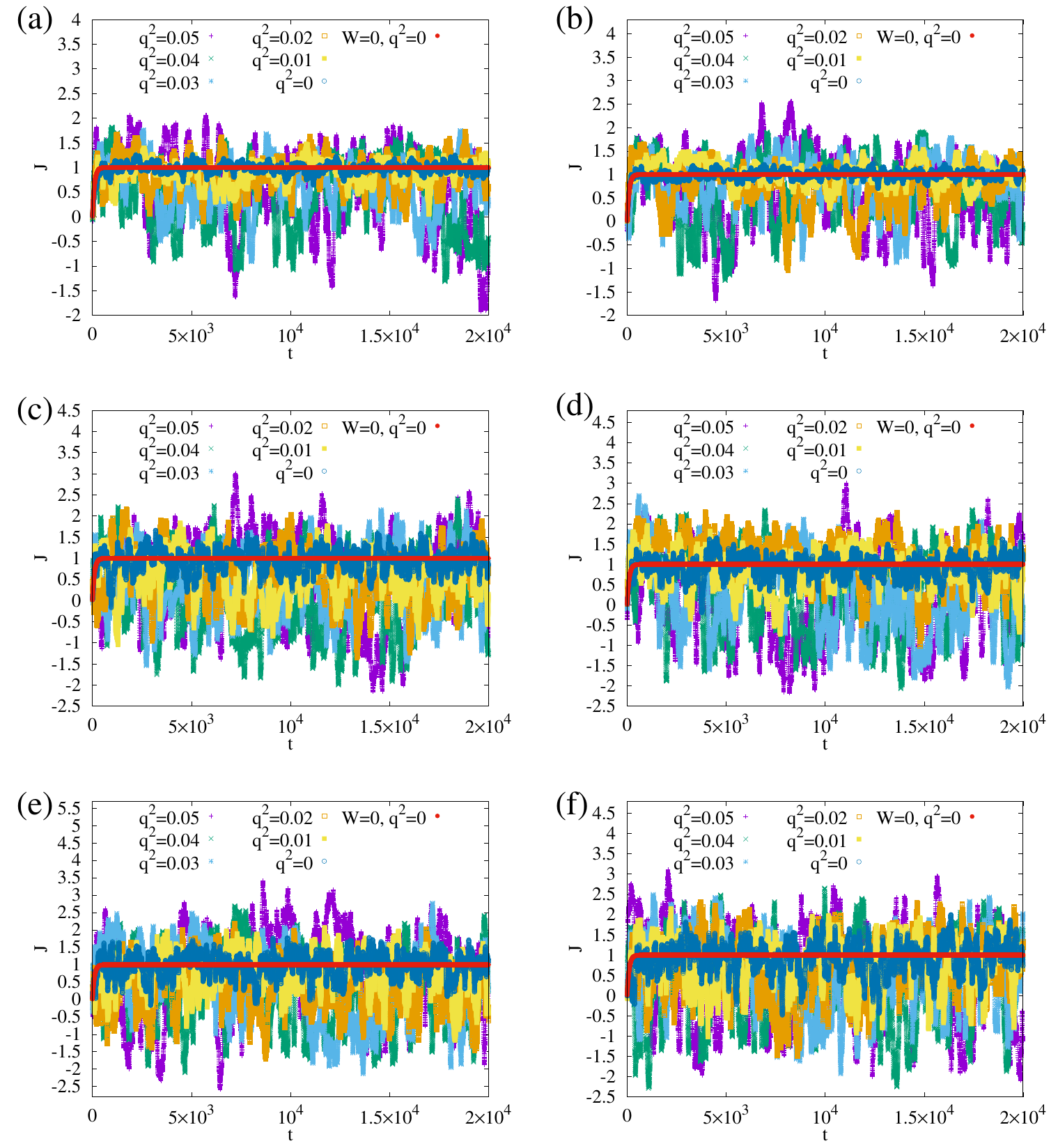}
\caption{Time evolution of the current $J$ of a single particle for different values of $q^2$ and $W=0.1$ (a,b), $W=0.3$ (c,d), $W=0.4$ (e,f).
These ones are compared to the classical unperturbed case, where $q^2=0$ and $W=0$.  The left column shows systems whose absolute value of the time averages is minimum among all realizations, while in the right column the time average is maximum. On a general basis, increasing $q^2$ and $W$ enhances the amplitude of the fluctuations, with a progressive shift towards values considerably lower than $1$. 
}\label{figS1}
\end{figure*}

\begin{figure*}[htbp]
\centering
\includegraphics[width=0.9\textwidth]{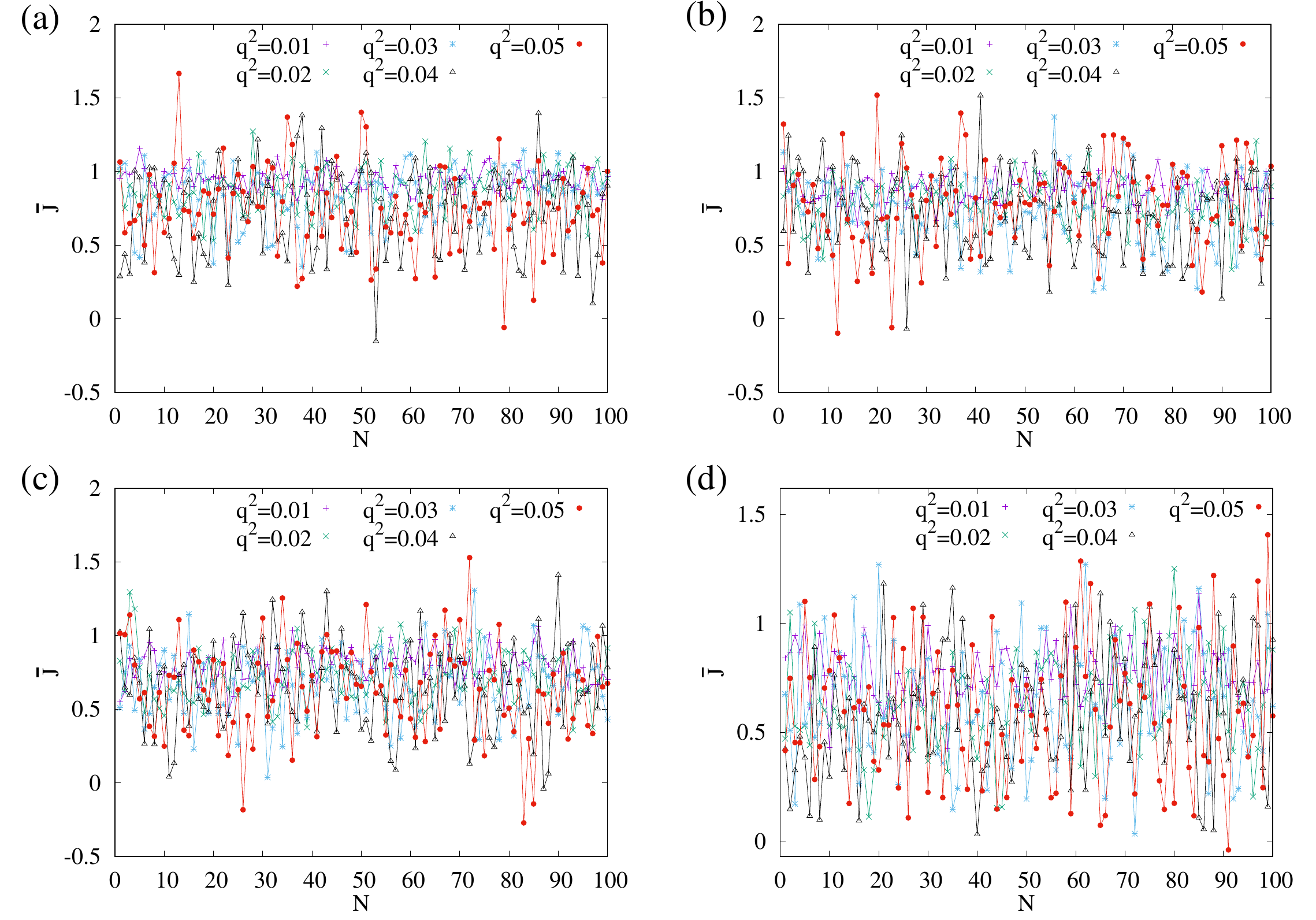}
\caption{Evolution of the time-averaged current $\bar{J}$ {\it vs} $N$, for $W=0.1$ (a), $W=0.2$ (b), $W=0.3$ (c), $W=0.4$ (d) and different values of $q^2$.  At low values of $W$, increasing $q^2$ yields mild effects on the amplitude of the fluctuations, while for larger values of $W$, the time-averaged current $\bar{J}$ displays larger perturbations with a gradual shift towards values   lower than $1$. 
}\label{figS2}
\end{figure*}

In Fig.\ref{figS1} we show the typical time evolution of $J$ for $W=0.1$ (a,b), $W=0.3$ (c,d) and $W=0.4$ (e,f), and different values of $q^2$.   
The left column (a,c,e) shows cases whose absolute value of the time average (over all realizations $N=100$) is minimum, while in the left one (b,d,f) the time average is maximum. Although they exhibit a behavior similar to that of Fig.2 of the main text, here we further note that
increasing $W$ and $q^2$ leads to larger amplitude fluctuations, progressively skewed towards values lower than $1$, which represents the unperturbed ($W=0$) classical ($q^2=0$) case. Also, for specific values of $q^2$, $J$ exhibits fluctuations of comparable amplitude, regardless
of whether its mean is at a minimum  (Fig.\ref{figS1} left column) or a maximum (Fig.\ref{figS1} right column).
These results complement the ones shown in Fig.2 of  the main text and, once again, support the view that large values of $W$ and $q^2$ can induce a current reduction. 
This is further corroborated in Fig.\ref{figS2},  where we show the evolution of $\bar{J}$ {\it versus} $N$, for $W=0.1$ (a), $W=0.2$ (b), $W=0.3$ (c), $W=0.4$ (d) and for increasing values of $q^2$. 
While for low values of $W$ the amplitude of the fluctuations remains basically confined between $0.5$ and $1$ with an average close to $1$, at large $W$ the current shows a marked skew towards values considerably lower than $1$, with significant fluctuations persistently spanning values between $0$ and $1$. 

\begin{figure*}[htbp]
\centering
\includegraphics[width=1.0\textwidth]{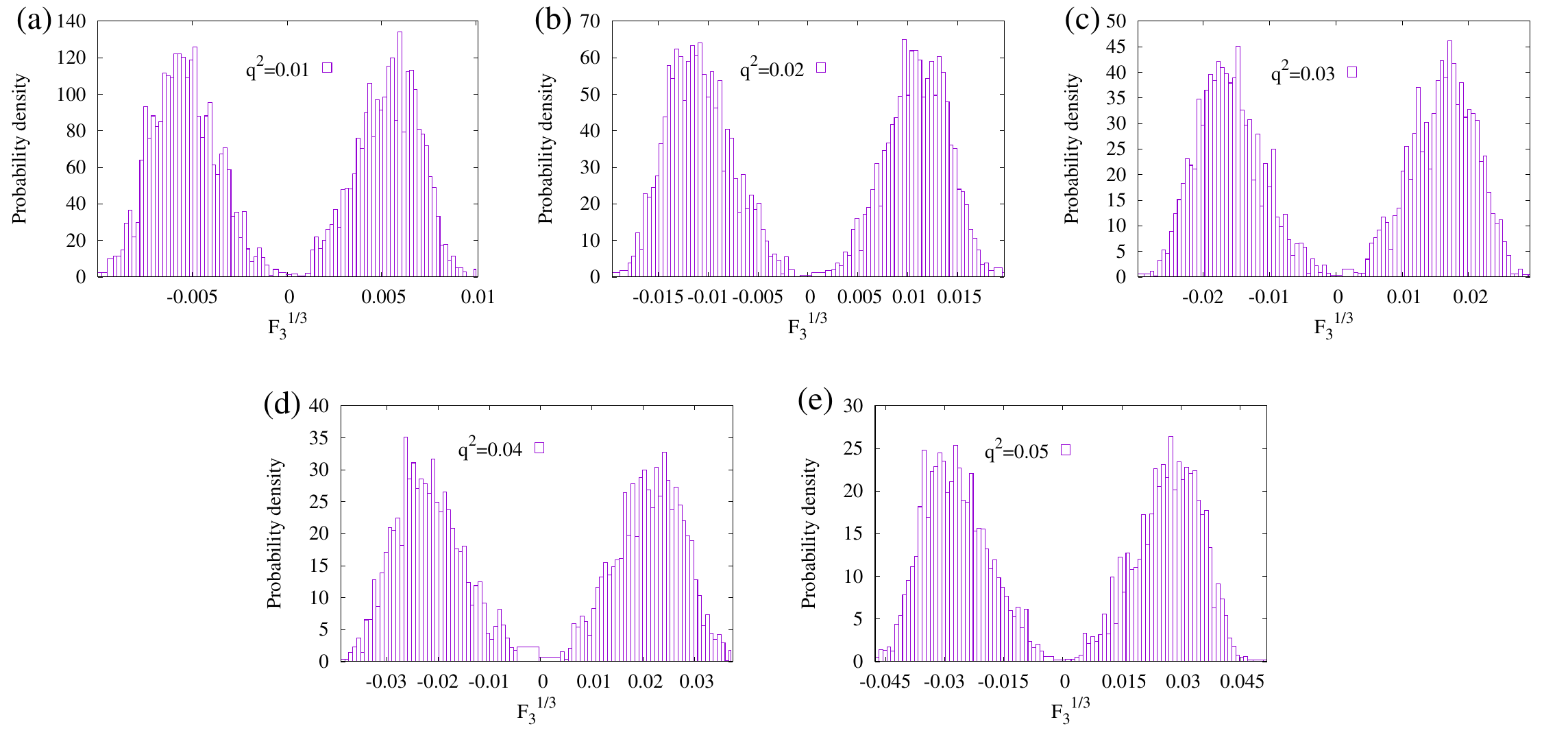}
\caption{
The distribution of $F_3^{1/3}$ displays a bimodal structure.
As $q^2$ increases, it broadens, extending toward larger positive and negative values since higher values of $F_3^{1/3}$ are attained.
}\label{figS3}
\end{figure*}

In Fig.\ref{figS3} we show the probability density function of $F_3^{1/3}$ (which appears in Eq.6 and 7 of the main text) for $W=0.2$ and different values of $q^2$.  The distribution exhibits a clear bimodal structure where positive and negative values rapidly alternate over time. This induces the negatively skewed distributions of $F_{1/3}$, since the kernel of  $F_{1/3}$ (see Eq.7 of the main text) weights recent events heavier than older ones, thus breaking the symmetry between positive and negative values. Note  additionally that increasing $q^2$ broadens the distribution because large values of the quantum force are attained. 

\end{document}